\documentclass[aps,prl,twocolumn,superscriptaddress]{revtex4}

\usepackage{graphicx}
\usepackage{color}
\usepackage[german,english]{babel}
\newcommand{\bk}{\ensuremath{\mathbf{k}}}

\begin{document}

\title{Probing of valley polarization in graphene via optical second-harmonic generation}

\author{T. O. Wehling}
\email{wehling@itp.uni-bremen.de}
\affiliation{Institut f{\"u}r Theoretische Physik, Universit{\"a}t Bremen, Otto-Hahn-Allee 1, 28359 Bremen, Germany}
\affiliation{Bremen Center for Computational Materials Science, Universit{\"a}t Bremen, Am Fallturm 1a, 28359 Bremen, Germany}
\author{A. Huber}
\affiliation{I. Institut f{\"u}r Theoretische Physik, Universit{\"a}t Hamburg, Jungiusstra{\ss}e 9, D-20355 Hamburg, Germany}
\author{A. I. Lichtenstein}
\affiliation{I. Institut f{\"u}r Theoretische Physik,
Universit{\"a}t Hamburg, Jungiusstra{\ss}e 9, D-20355 Hamburg,
Germany}
\author{M. I. Katsnelson}
\affiliation{Radboud University of Nijmegen, Institute for
Molecules and Materials, Heijendaalseweg 135, 6525 AJ Nijmegen,
The Netherlands}


\date{\today}

\begin{abstract}
Valley polarization in graphene breaks inversion symmetry and therefore leads to second-harmonic generation. We present a complete theory of this effect within a single-particle approximation. It is shown that this may be a sensitive tool to measure the valley polarization created, e.g., by polarized light and, thus, can be used for a development of ultrafast valleytronics in graphene.
\end{abstract}
\maketitle

\begin{figure}%
\includegraphics[width=\columnwidth]{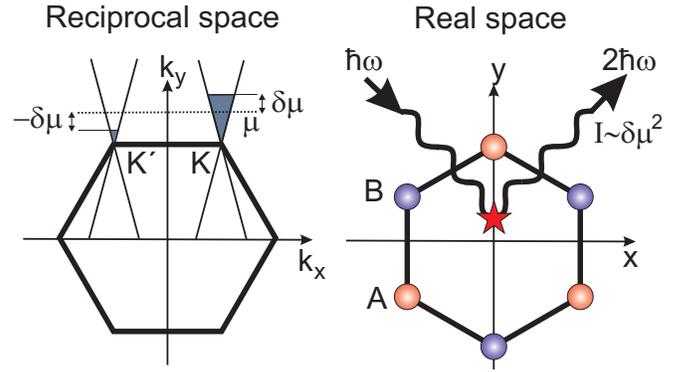}%
\caption{Illustration of second harmonic generation in graphene. SHG requires breaking of inversion symmetry which can be achieved through valley polarization as illustrated in the left panel. Valley polarization is modelled in terms of different chemical potentials $\mu\pm\delta\mu$ in valley K and K'. With the choice of the coordinate system illustrated in the right panel, valley polarization breaks the $x\to -x$ mirror symmetry. The second harmonic intensity $I$ is proportional to $\delta\mu^2$.}%
\label{fig:exp_scheme}%
\end{figure}

The unique electronic properties of graphene \cite{GK07,CNetal08,K12} open ways for many interesting and unusual applications. In particular, a concept of {\it valleytronics} was suggested \cite{RTB07}, that is, a manipulation of valley degree of freedom (conical points K and K'), in analogy with the well-known field of spintronics \cite{ZFD04}. Up to now, many different ways for the creation of the valley polarization in graphene have been proposed (see, e.g., Refs. \onlinecite{MBM08,GTEM11,Yetal13}). At the same time, {\it detection} of the valley polarization is a tricky issue. The first suggestion, the use of a superconducting current through graphene \cite{AB07} does not look suitable for practical applications, e.g., due to a requirement of low temperatures. It was mentioned in Ref. \onlinecite{GTEM11} that the breaking of inversion symmetry by the valley polarization can be probed via second-harmonic generation (SHG), a well-known nonlinear optical effect \cite{B92}. Together with their suggestion to use circularly polarized light to create the valley polarization (recently, it was experimentally realized for another two-dimensional crystal, MoS$_2$ \cite{MHSH12}) it would open a way to ultrafast valleytronics where all manipulations with the valley degree of freedom are performed via short laser pulses, as illustrated in Fig. \ref{fig:exp_scheme}. In spintronics, this is now one of the most prospective lines of development \cite{KKR10}.

There is, however, a problem. SHG is related to the term in the current density $\vec{j}$ proportional to the square of the electric field $\vec{E}$:
\begin{equation}
j_{\alpha} = \chi_{\alpha \beta \gamma} E_{\beta} E_{\gamma}.
\label{nonlin1}
\end{equation}
For a system with inversion center $\hat{\chi} =0$. This does not mean, however, that SHG is impossible since the photon wave vector $\vec{q}$ plays the role of a factor violating inversion symmetry, and there is a contribution to the current
\begin{equation}
j_{\alpha} = \phi_{\alpha \beta \gamma \delta} E_{\beta} E_{\gamma} q_{\delta}.
\label{nonlin2}
\end{equation}
For the case of graphene and other two-dimensional electron systems it was calculated in Ref. \onlinecite{M11}. In comparison with Eq.(\ref{nonlin1}) it contains a relativistic smallness parameter $v_F/c \approx 1/300$ where $v_F$ is the Fermi velocity and $c$ is the velocity of light. At the same time, the current in Eq.(\ref{nonlin1}) is expected to be proportional to the valley polarization. The order of magnitude of the valley polarization which can be really probed via SHG depends on explicit values of the tensor $\hat{\chi}$, which will be calculated in this work.

We start with a derivation of the effective Hamiltonian of electron-photon interaction for the case of graphene (c.f. e.g. Refs. \onlinecite{SPG08,K12}), as the case of nonlinear optics requires a special care. Let us consider a general Hamiltonian of band electrons in electromagnetic field described by the vector potential $\mathbf{A}(\mathbf{r},t)$:
\begin{equation}
H=\sum_{ij,LL',\sigma}t_{ij}^{L'L}  \exp\left(i\frac{e}{c}\int_{\mathbf{R}_{jL'}}^{\mathbf{R}_{iL}}d\mathbf{r}\mathbf{A}(\mathbf{r},t)\right)   c^{\dagger}_{iL\sigma}c_{jL'\sigma}
\label{ham}
\end{equation}
where $\mathbf{R}_{iL}$ is the atomic position and $L=(n,l,m,\gamma)$ is a combined index of quantum numbers of atom $\gamma$ (in the equations we assume $\hbar =1$). The atomic positions can be separated into two parts
\begin{equation}
\mathbf{R}_{iL}=\mathbf{R}_{i}+\rho_{L},
\label{R}
\end{equation}
where the former indexes the unit cell $i$ and the latter the atom $L$ within the cell in the case of a multi-atomic unit cell (like the honeycomb lattice of graphene). We assume, as usual, that the interaction with the electromagnetic field is taken into account via Peierls substitution
\begin{equation}
c_{iL\sigma}^{\dagger}\rightarrow c_{iL\sigma}^{\dagger}\exp\left(i\frac{e}{c}\int^{\mathbf{R}_{iL}}d\mathbf{r}\mathbf{A}(\mathbf{r},t)\right)
\label{peiers}
\end{equation}
for the electron creation operators $c_{iL\sigma}^{\dagger}$ and similar for the electron annihilation operators $c_{iL\sigma}$. $t_{ij}^{L'L}$ are the parameters of the band-structure Hamiltonian.

Since we are interested in terms up to second order in the vector potential we expand the hopping and treat the additional terms proportional to the vector potential as perturbation. The Hamiltonian becomes then
\begin{widetext}
\begin{eqnarray}
 &H& \equiv H^{(0)}+H^{(1)}+H^{(2)}+\mathcal{O}(\mathbf{A}^{3}) \label{theory}
\\
  &&  =\sum_{ij,LL',\sigma}t_{ij}^{L'L} c^{\dagger}_{iL'\sigma}c_{jL\sigma}  + i\frac{e}{c}A_{\alpha}(t)\sum_{ij,LL',\sigma}t_{ij}^{L'L}(R_{iL'\alpha}-R_{jL\alpha})c^{\dagger}_{iL'\sigma}c_{jL\sigma} \nonumber \\
  && \ \ \ \  +\frac{1}{2}(i\frac{e}{c})^{2}A_{\alpha}(t)A_{\beta}(t)\sum_{ij,LL',\sigma}t_{ij}^{L'L}(R_{iL'\alpha}-R_{jL\alpha})(R_{iL'\beta}-R_{jL\beta}) c^{\dagger}_{iL'\sigma}c_{jL\sigma}+\mathcal{O}(\mathbf{A}^{3}) \nonumber
\end{eqnarray}
\end{widetext}
where the second equation is defined in powers of the vector potential and we further assumed that the vector potential slowly varies in $\mathbf{r}$. 
With a basis transformation to Bloch waves $c_{\bk L\sigma}=\frac{1}{\sqrt{N}}\sum_j\exp\left(i\bk\mathbf{R}_{j}\right)c_{jL\sigma}$ the bare part of the Hamiltonian is diagonalized according to $H^{(0)}=\sum_{\bk,LL',\sigma} H^{0}_{\mathbf{k},LL'}c^{\dagger}_{\mathbf{k}L\sigma}c_{\mathbf{k}L'\sigma}$ with
\begin{equation}
H^{0}_{\mathbf{k},LL'}= \sum_{ij}t_{ij}^{L'L}\exp(-i\mathbf{k}(\mathbf{R}_{i}-\mathbf{R}_{j})).
\label{hk}
\end{equation}
Now one can distinguish between two currents which are defined using the Hamiltonian (\ref{theory}) by
\begin{equation}
j^{(1)}_{\alpha}\equiv \left.\frac{\delta H}{\delta A_{\alpha}(\mathbf{r},t)} \right|_{\textbf{A}=0} =e\sum_{LL',\mathbf{k},\sigma}v_{\mathbf{k}\alpha}^{LL'}c^{\dagger}_{\mathbf{k}L'\sigma}c_{\mathbf{k}L\sigma}
\label{j1}
\end{equation}
\begin{equation}
j^{(2)}_{\alpha\beta}\equiv \left. \frac{\delta^{2} H}{\delta A_{\alpha}(\mathbf{r},t) \delta A_{\beta}(\mathbf{r},t)} \right|_{\textbf{A}=0} =e^{2}\sum_{LL',\mathbf{k},\sigma}v_{\mathbf{k}\alpha\beta}^{LL'}c^{\dagger}_{\mathbf{k}L'\sigma}c_{\mathbf{k}L\sigma},
\label{j2}
\end{equation}
where
\begin{widetext}
\begin{equation}
v_{\mathbf{k}\alpha}^{LL'}\equiv i\sum_{i-j}t_{ij}^{L'L}(R_{iL'\alpha}-R_{jL\alpha})\exp\left(-i\mathbf{k}(\mathbf{R}_{i}-\mathbf{R}_{j})\right)
\label{fermivelocity}
\end{equation}
and
\begin{equation}
v_{\mathbf{k}\alpha\beta}^{LL'}\equiv (i)^{2}\sum_{i-j}t_{ij}^{L'L}(R_{iL'\alpha}-R_{jL\alpha})(R_{iL'\beta}-R_{jL\beta})\exp\left(-i\mathbf{k}(\mathbf{R}_{i}-\mathbf{R}_{j})\right).
\label{fermivelocity2}
\end{equation}
With Fourier transform of the band Hamiltonian (\ref{theory})
we can reexpress the generalized velocitites resulting from Eq. (\ref{fermivelocity}) according to
\begin{equation}
v_{\mathbf{k}\alpha}^{LL'}=\left(\partial_{k_{\alpha}} -i(\rho^{\alpha}_{L'}-\rho_{L}^{\alpha})\right)H^{(0)}_{\mathbf{k},LL'}
\end{equation}
and Eq.(\ref{fermivelocity2}) leads to
\begin{equation}
v_{\mathbf{k}\alpha\beta}^{LL'}=\left(   \partial_{k_{\alpha}}\partial_{k_{\beta}} +i(\rho^{\alpha}_{L'}-\rho_{L}^{\alpha})\partial_{k_{\beta}} +i(\rho^{\beta}_{L'}-\rho_{L}^{\beta})\partial_{k_{\alpha}} -(\rho^{\alpha}_{L'}-\rho_{L}^{\alpha})(\rho^{\beta}_{L'}-\rho_{L}^{\beta})\right)H^{(0)}_{\mathbf{k},LL'}.
\end{equation}
\end{widetext}

We will use these general expressions for the particular case of graphene, in a single-band approximation ($\pi$-bands only) taking into account only the nearest-neighbor ($t$) and the next-nearest-neighbor ($t'$) hopping parameters \cite{K13}; the latter can be important since it breaks the electron-hole symmetry of the Hamiltonian which as we will see is essential for SHG.

\begin{figure}
	\centering
		\includegraphics[width=0.9\columnwidth]{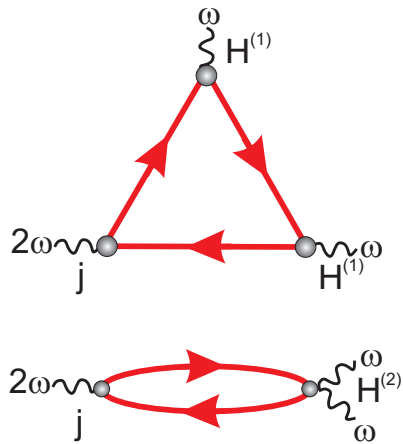}
		\caption{Two second-order contributions to the non-linear susceptibility. Top: triangle diagram, bottom: unlabeled non-linear bubble diagram. Solid lines are electron Green's functions and wavy tails indicate photons involved in the processes.}
\label{diag}
\end{figure}

There are two contributions to the electric current quadratic in the vector potential $\mathbf{A}(\mathbf{r},t)$. Note that we now switch to the response of an electric field by using the identity
\begin{equation}
\frac{1}{c}A_{\alpha}(\textbf{r},t)=-i\frac{E_{\alpha}(\textbf{r},t)}{\omega}.
\label{michi}
\end{equation}
The contributions to the nonlinear optical conductivity (\ref{nonlin1}) via Feynman diagrams are drawn in Fig. \ref{diag}.

\begin{figure}
	\centering
		\includegraphics[width=0.99\columnwidth]{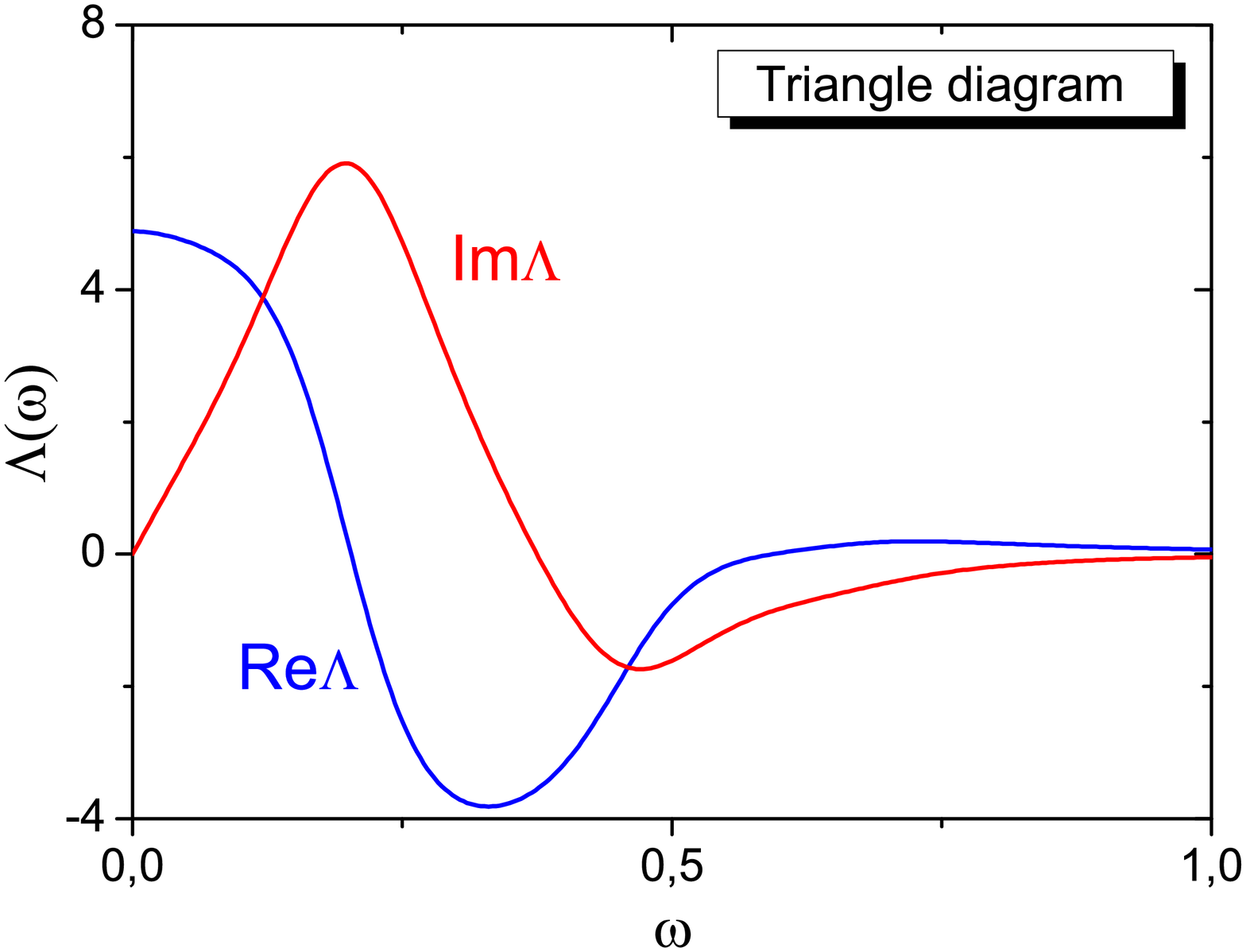}
		\includegraphics[width=0.99\columnwidth]{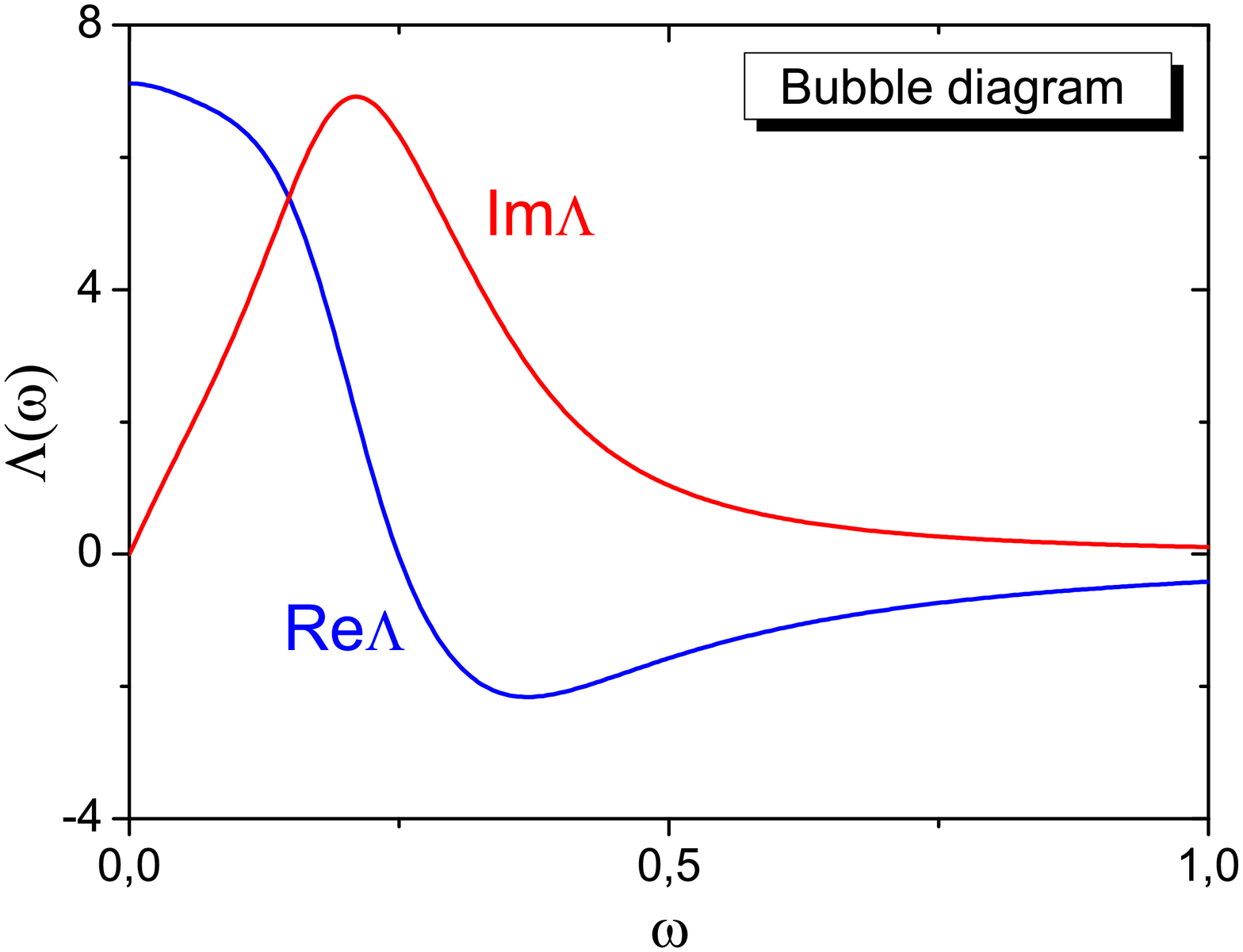}
		\caption{Computational results for $\Lambda=-\left( \hbar \omega \right)^2\frac{\partial \chi_{xxx}}{\partial \mu}$ (in the units of $e^3 a/\hbar$, $a$ is the lattice constant) as a function of real frequency $\omega$ (in the units of $t/\hbar$). The total answer is the sum of the triangle and bubble contributions.}
\label{bubbletriangle}
\end{figure}
\begin{widetext}
The corresponding algebraic equation for the triangle diagram is given by
\begin{eqnarray}
\chi^{triangle} _{\alpha\beta\gamma}(i\omega ,i\omega ,2i\omega )&=&-i\frac{e^{3}}{\omega ^{2}}%
\frac{1}{\beta}\sum_{\nu }\sum_{\mathbf{k}}\sum_{L_1 \ldots L_6}v^{L_6L_1}_{\textbf{k}\alpha}
G_{L_1L_2}(\mathbf{k,}i\nu)
\nonumber \\
&&
\cdot v^{L_2L_3}_{\mathbf{k}\beta}%
G_{L_3L_4}(\mathbf{k,}i\nu + i\omega )
v^{L_4L_5}_{\mathbf{k}\gamma}G_{L_5L_6}(\mathbf{k,}%
i\nu -i\omega ) \label{sigma3tri}
\end{eqnarray}
and for the non-linear bubble diagram
\begin{eqnarray}
&&\chi^{bubble} _{\alpha\beta\gamma}(i\omega ,i\omega ,2i\omega ) = -i\frac{e^{3}}{\omega ^{2}}%
\frac{1}{\beta}\sum_{\nu }\sum_{\mathbf{k}}\sum_{L_1 \ldots L_4}v^{L_4L_1}_{k\alpha\beta}
G_{L_1L_2}(\mathbf{k,}i\nu - i\omega)
v^{L_2L_3}_{\mathbf{k}\gamma}%
G_{L_3L_4}(\mathbf{k,}i\nu + i\omega ).
 \label{sigma3bub}
\end{eqnarray}
\end{widetext}
Here $L_1,...,L_6$ are pseudospin indices, $\beta=1/T$ is the inverse temperature (we use the units $\hbar = k_B=1$) and
\begin{equation}
\widehat{G}\left( i\nu \right) =\frac 1{i\nu +\mu -\widehat{H}}
\end{equation}
is the Green's function with $\mu$ being the chemical potential counted from the neutrality (conical) point. Thus, the non-linear susceptibility is given by
\begin{equation}
\chi_{\alpha\beta\gamma}=\chi^{bubble} _{\alpha\beta\gamma}+\chi^{triangle} _{\alpha\beta\gamma}.
\label{sigma3}
\end{equation}
Note that the minus sign from the fermion loop should be taken into account in both diagrams. The factor $\frac{1}{\omega^{2}}$ appears due to Eq.~(\ref{michi}). We pass, as usual \cite{M00} to imaginary (Matsubara) frequencies; at the end of the calculations the anaytical continuation to the real axis $i\omega \rightarrow \omega + i \delta$ is performed.

If we take into account electron-electron interactions the nonlinear conductivity will be renormalized by three-leg and six-leg electron vertices; the corresponding expressions can be found in Ref. \onlinecite{KL10}.

It is obvious by inversion symmetry that for the non-valley-polarized case $\hat{\chi} =0$. We mimic the valley polarization by splitting the Brillouin zone into two symmetrically chosen parts, one containing the point K and the other part containing the point K', and assuming different chemical potentials for these two parts. We expand then all the quantities dependent on the chemical potential as
\begin{equation}
f(\mu+\delta\mu)-f(\mu-\delta\mu)\approx 2\frac{\partial f(\mu)}{\partial \mu} \delta\mu
\end{equation}
We evaluated the derivative $\partial \chi/\partial \delta\mu$ analytically using Eqs. (\ref{sigma3tri}), (\ref{sigma3bub}) and then performed a numerical summation over Matsubara frequencies and wave vectors involving half of the Brillouin zone. We choose $\beta=40$/eV, which corresponds to a temperature of 290 K. In this case, sampling of the Brillouin with $121\times 121$ k-points and summation of 200 (1000) fermionic Matsubara frequencies are required to reach convergence for the triangle (non-linear bubble) diagram at bosonic Matsubara frequencies $\Omega_n=2\pi n/\beta$ in the range of $n=1,..,20$.

A symmetry analysis shows that there are only two independent components of the tensor $\hat{\chi}$, $\chi_{xxx}=-\chi_{xyy}=-\chi_{yxy}=-\chi_{yyx}$ and $\chi_{yyy}=-\chi_{yxx}=-\chi_{xxy}=-\chi_{xyx}$ \cite{B92}. With the choice of coordinates made here, the K and K' point of the Brillouin zone are on the positive / negative x-axis, see Fig. \ref{fig:exp_scheme}. Thus valley polarization breaks inversion symmetry with respect to the x-direction, $x\to -x$, but the $y\to -y$ symmetry is preserved. Thus, we have $\chi_{yyy}=0$ and we will show the results only for $\chi_{xxx}$.

The computational results for the case of finite chemical potential $\mu=0.2t$ are shown in Fig. \ref{bubbletriangle}. For the case $\mu=0,t'=0$ one finds $d\chi_{xxx}/d\mu=0$, due to electron-hole symmetry. Nearest neighbor-hopping $t'$ breaks this symmetry and leads to non-zero $d\chi_{xxx}/d\mu$ even at $\mu = 0$ and according to Ref. \onlinecite{K13} we have $t'\approx 0.1t$. Our calculations show, however, that for  $\mu=0.2t$ the effects of finite $t'$ are negligible leading only to a few-percent corrections.

One can see from Fig. \ref{bubbletriangle} that a dimensionless quantity $\Lambda$ characterizing the valley polarization induced SHG is pretty large, of the order of ten, despite the smallness of the ratios $t'/t$ and  $\mu/t$.  It is consistent with the computational results \cite{Getal04,Zetal06,PP09} on SHG in chiral nanotubes which turned out to be strongly enhanced in comparison with conventional materials without inversion symmetry. A comparison with the results of Ref. \onlinecite{M11} shows that the valley-polarization induced SHG will be dominant if $|\delta \mu|/t > 0.01 v_F/c \approx 3 \cdot 10^{-5}$.
Note, that one additional smallness in order of magnitude originates from the factor $3/8\pi \approx 0.1$ in Ref. \onlinecite{M11} and another
one from the fact that  $\Lambda \approx 10$.

Typical nonlinear crystals have second order nonlinear susceptibilities on the order of $\tilde\chi=0.1$ to $100~$pm/V \cite{B92}. It is interesting to see which amount of valley polarization is required to reach this order in graphene. From the current density $j=\chi E^2$ we obtain the oscillating in-plane charge density $|\sigma|=|j|/c$ and the associated electric field $E_{2\omega}=\sigma/\epsilon_0=\chi E^2/\epsilon_0 c=\tilde \chi E^2$. With photon energies on the order of $\hbar\omega=1.5$~eV$\approx 0.5t$ and $\Lambda\approx 10$ we find thus $\tilde \chi/\delta \mu=-\Lambda(e^3a/\hbar)/(\epsilon_0 c(\hbar \omega)^2)\approx 1$(\AA/V)/eV$=100$~(pm/V)/ eV. Thus $\delta\mu\gtrsim 1$ meV is required to reach $\tilde\chi=0.1$ pm/V. 

This means that SHG is, indeed, a very efficient tool to probe the valley polarization in graphene.
Our results show that ``triangle'' and ``bubble'' contributions to the second-harmonic generation are in general comparable.
Also, one can see that they have quite a similar frequency dependence. An alternative way to probe the valley polarization is rectification, that is, generation of dc current under laser pulses. This process is described by the quantity $\chi_{\alpha\beta\gamma}(\omega ,\omega ,0)$ which is of the same order of magnitude as $\chi_{\alpha\beta\gamma}(\omega ,\omega ,2\omega )$ calculated here. It would be very interesting to probe both of these effects experimentally in graphene with valley polarization.

\section*{Acknowledgements}
The authors acknowledge financial support from European Union Seventh Framework Program under grant agreement Graphene Flagship and the German Research Foundation (DFG) through SPP 1459. M.I.K. also thanks financial support from the European Research Council Advanced Grant program (contract 338957).

\end{document}